\documentclass[a4paper, fontsize = 10pt, DIV=calc, pdflatex]{scrartcl}
\usepackage{graphicx, nicefrac, siunitx}
\usepackage[hyperfootnotes=false]{hyperref}
\usepackage{graphics}
\usepackage{times}
\usepackage{csquotes, xpatch}
\usepackage{amsmath, amssymb}
\usepackage{setspace}
\usepackage{color}
\usepackage{xcolor}
\usepackage[font = small, format = plain, labelfont = bf]{caption}
\usepackage[colaction]{multicol}
\usepackage{float}
\usepackage{tabularx}

\usepackage[a4paper, top=8mm, tmargin=8mm, left=13mm, right=15mm, textwidth=182mm, bottom=10mm, bmargin=20mm,]{geometry}
\usepackage{changepage }
\renewcommand{\footnoterule}{%
  \kern -3pt
  \hrule width \textwidth height 0.5pt
  \kern 2pt
}
\begin{document}
\newcommand\Tstrut{\rule{0pt}{2.6ex}}         
\newcommand\Bstrut{\rule[-0.9ex]{0pt}{0pt}}   

\newcommand{\fr}[1]{\textbf{\textcolor{red}{#1}}} 
\newcommand{\hci}[3]{\textsuperscript{#1}#2\textsuperscript{#3}} 
\newcommand{\Sec}[1]{ \textbf{#1}\newline}

\newcommand\multicollinenumbers{%
 \linenumbers
 \def\makeLineNumber{\docolaction{\makeLineNumberLeft}{}{\makeLineNumberRight}}}

\onecolumn
\normalsize
\begin{flushleft}
    \begin{flushleft}
        \fontsize{20}{12}
        \hrulefill
    \end{flushleft}
    \fontsize{30}{12}
    \vspace{5pt}
    \textbf{Stringent test of QED with\\ hydrogenlike tin\\}
    \vspace{30pt}

\end{flushleft}
\begin{flushleft}
    \begin{adjustwidth}{10pt}{0pt}
        \begin{flushleft}
                \fontsize{10}{12}
                \textbf{J. Morgner\footnote{Max Planck Institute for Nuclear Physics, 69117 Heidelberg, Germany, Helmholtz-Institut, \textsuperscript{2}GSI Helmholtzzentrum für Schwerionenforschung, Mainz 55128, Germany, \textsuperscript{$\dagger$} jonathan.morgner@mpi-hd.mpg.de, alphatrap@mpi-hd.mpg.de}\textsuperscript{,$\dagger$}, B. Tu\textsuperscript{1}, C. M. König\textsuperscript{1}, T. Sailer\textsuperscript{1}, F. Heiße\textsuperscript{1}, H. Bekker\textsuperscript{2}, B. Sikora\textsuperscript{1}, C. Lyu\textsuperscript{1}, V. A. Yerokhin\textsuperscript{1}, Z. Harman\textsuperscript{1}, J. R. Crespo López-Urrutia\textsuperscript{1}, C. H. Keitel\textsuperscript{1}, S. Sturm\textsuperscript{1}, and K. Blaum\textsuperscript{1}\\}
        \end{flushleft}
        \vspace{20pt}
            \setstretch{1.2}
            Inner-shell electrons naturally sense the electric field close to the nucleus, which can reach extreme values beyond \SI{e+15}{\volt/\cm} for the innermost electrons~\cite{beier_g_j_2000}.
            Especially in few-electron highly charged ions, the interaction with the electromagnetic fields can be accurately calculated within quantum electrodynamics (QED), rendering these ions good candidates to test the validity of QED in strong fields.
            Consequently, their Lamb shifts were intensively studied in the last decades~\cite{shabaev_stringent_2018, beiersdorfer_testing_2010}.
            Another approach is the measurement of \textit{g} factors in highly charged ions~\cite{verdu_electronic_2004, hannen_lifetimes_2019, sturm_g_2011, kohler_isotope_2016}.
            However, so far, either experimental accuracy or small field strength in low-\textit{Z} ions~\cite{hannen_lifetimes_2019, sturm_g_2011} limited the stringency of these QED tests.
            Here, we report on our high-precision, high-field test of QED in hydrogenlike \textsuperscript{118}Sn\textsuperscript{49+}.
            The highly charged ions were produced with the Heidelberg-EBIT~\cite{martinez_heidelberg_2007} (electron beam ion trap) and injected into the \textsc{Alphatrap} Penning-trap setup~\cite{sturm_alphatrap_2019}, where the bound-electron \textit{g} factor was measured with a precision of 0.5 parts-per-billion.
            For comparison, we present state-of-the-art theory calculations, which together test the underlying QED to about \SI{0.012}{\percent}, yielding a stringent test in the strong-field regime.
            With this measurement, we challenge the best tests via the Lamb shift and, with anticipated advances in the \textit{g}-factor theory, surpass them by more than an order of magnitude.
    \end{adjustwidth}
\end{flushleft}
\vspace{10pt}
\hrule
\begin{multicols}{2}

In 1963 Richard Feynman called quantum electrodynamics (QED) the greatest success in the physical sciences~\cite{feynman_feynman_2011}.
Describing the ubiquitous interactions of charges and the electromagnetic field with real and virtual photons, QED is the prime example of quantum field theories.
Experimentally, QED has been tested with high stringency in low electromagnetic fields.
Such tests are closely related to the determination of fundamental constant, as e.g. the recent measurement of the $g-2$ value, which allowed to extract the fine-structure constant $\alpha$ with a precision of \SI{1.1e-10}{}~\cite{fan_measurement_2023}.
In contrast only few experimental tests have been carried out at high electromagnetic field strengths.
Here, bound-state QED can yield high accuracy in the prediction of atomic and molecular systems.
Thus, testing QED calculations still has wide implications for many branches of science.\newline
In the past, muonic atoms have been studied extensively, leading to a series of stringent tests of the vacuum polarization in strong electric fields~\cite{dixit_new_1975, beltrami_new_1986, borie_energy_1982}.
Furthermore, Standard-Model predictions of muonic fine-structure splittings are inconsistent with experimental data~\cite{valuev_evidence_2022, haga_full-relativistic_2005, haga_reanalysis_2007}.
Also recently, the muon $(g-2)$ value has been remeasured, and shows a 4.2-$\sigma$ discrepancy~\cite{muon_g-2_collaboration_measurement_2021}.
As a consequence, this strongly motivates further tests of QED in strong electromagnetic fields.

Highly charged ions are an interesting candidate for such tests as due to the strong interaction between the (few) electrons and the nucleus, these systems also show enhanced sensitivity for potential new physics~\cite{kozlov_highly_2018}.
In these few-electron systems, the electric field experienced by the remaining electrons can exceed \SI{e+15}{\volt/\cm}~\cite{beier_g_j_2000}, hence the electronic wave function is perturbed strongly, resulting in modified properties, that can be measured and compared to theoretical predictions.
Thus far, bound-state QED in high-$Z$ highly charged ions has been probed most accurately by measurements of the Lamb shift~\cite{beiersdorfer_measurement_2005, gumberidze_quantum_2005}.
Presently, calculations of the Lamb shift employ an 'all-order' approach including all QED effects in one- and two-loop Feynman diagrams~\cite{yerokhin_lamb_2015}.
For testing bound-state QED using the magnetic moment or the gyromagnetic factor ($g$ factor) of the bound electron the theoretical approach is similar.
Due to the additional interaction with a magnetic field, its calculation requires the inclusion of additional terms. 
But different to the Lamb shift, the calculation of the $g$ factor two-loop contributions with an all-order approach is not yet completed.
Therefore these are calculated using a series expansion in $Z\alpha$, which is expected to have large uncertainty at high-$Z$, due to the strong scaling with $Z$.
Here, $Z$ is the atomic number, and $\alpha$ the fine-structure constant.
In low-$Z$ systems, as the expansion coefficient $Z\alpha$ is small, high accuracy can be achieved in the prediction.
Many systems with different charge states have been probed in the past~\cite{sturm_g_2011, wagner_g_2013, kohler_isotope_2016, arapoglou_g_2019, sailer_measurement_2022}.
Furthermore, the measurement of the hydrogenlike carbon $g$ factor allowed to determine the electron mass to an unprecedented precision~\cite{sturm_high-precision_2014}.

\begin{figure*}[ht]
    \centering
    \includegraphics[width = \textwidth]{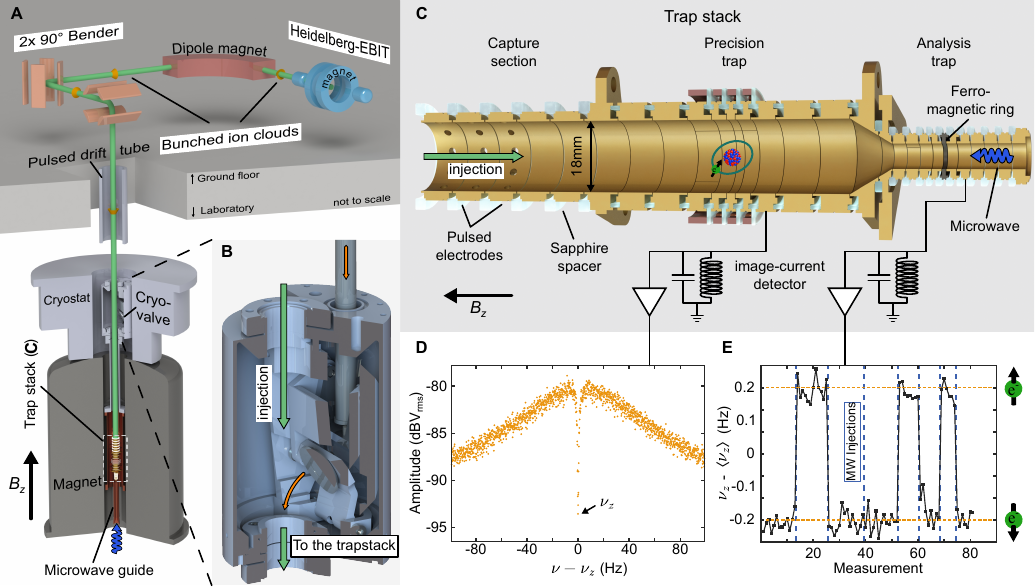}
    \caption{
    \textbf{Fig. 1: Experimental setup for production, trapping and detection of hydrogenlike \textsuperscript{118}Sn\textsuperscript{49+}.}
    The highly charged ions are produced in the Heidelberg-EBIT.
    Via a room temperature beamline, the ions are transported into the \textsc{Alphatrap} magnet as shown in \textbf{A}.
    \textbf{B}, the cryovalve allows to maintain an ultra-high vacuum within the trap chamber.
    Subfigure \textbf{C} shows the 'trap stack' of the experimental setup.
    The ions are captured in the capture section by pulsing the applied voltage at the moment the ions are in the trap.
    Below is the precision trap, a 7-electrode trap in which the frequency ratio $\Gamma_0 = \nu_L/\nu_c$ is measured.
    An image-current detector is used to detect the particle motion in the trap.
    The voltage applied to the centre electrode is around \SI{-59}{\volt}.
    On the bottom of the trap stack, the analysis trap is located, which has a strong magnetic bottle, allowing the detection of the spin state of the bound electron.
    \textbf{D}, Fourier spectrum of the image-current detector with a \textsuperscript{118}Sn\textsuperscript{49+} particle in resonance.
    Fitting this 'dip' gives the axial frequency of the particle.
    \textbf{E}, axial frequency change (about \SI{\sim 300}{\milli\hertz}) after flipping the electron spin by microwave irradiation at the Larmor frequency.
    }
    \label{fig:Trap}
\end{figure*}
The so far heaviest measured $g$ factor of hydrogenlike ions is \textsuperscript{28}Si\textsuperscript{13+} which allowed for a stringent test of QED in low to medium-$Z$ ions~\cite{sturm_g_2011, sturm_g-factor_2013}.\\
Here, we report on our high-precision $g$-factor measurement in hydrogenlike \textsuperscript{118}Sn\textsuperscript{49+}, reaching directly into the medium-to-high-$Z$ range.
To achieve this we produce the hydrogenlike ions externally in the Heidelberg-EBIT~(\cite{martinez_heidelberg_2007}) which can reach significantly higher charge states than the ion sources that were previously available for this type of measurement.
From there the ions are transported into the \textsc{Alphatrap} apparatus, where we capture them in order to perform high-precision spectroscopy of the bound-electron $g$ factor.
We further compare the measured value with its state-of-the-art theory prediction, which tests bound-state QED in a mean electric field of \SI{1.6e+15}{\volt/\cm}, 60 times stronger compared to the \textsuperscript{28}Si\textsuperscript{13+} measurement, the so far strongest field for a precise $g$-factor measurement.
\\
For the presented measurement, an enriched sample of Sn-118 was heated in an oven-source for injection into the Heidelberg EBIT~\cite{martinez_heidelberg_2007}.
In the electron beam ion trap (EBIT), a $200$-mA electron beam focused by a $7$-T magnetic field to a waist of a few ten \SI{}{\micro\m} crosses the atomic beam in the centre electrode.
With a kinetic energy of around \SI{45}{\kilo\eV}, well above the binding energy of the $K$ shell (\SI{\approx35}{\kilo\eV}~\cite{kramida_nist_2021}), electrons impacting on the tin atoms sequentially generate higher charge states until the charge-state distribution reaches a steady state.
For the production and extraction of hydrogenlike \textsuperscript{118}Sn\textsuperscript{49+}, a charge-breeding time of \SI{60}{\second} was used.
After this, a fast pulse on the central electrode ejects the trapped highly charged ions.
The ion bunch, with a kinetic energy around $\SI{7}{\kilo\eV} \times N_q$ ($N_q$ as the charge state), is transported through a room-temperature beamline, where the required charge state is separated with a dipole magnet.
A schematic view of the beamline can be seen in Fig.~\ref{fig:Trap}A.
Various ion-optical elements guide the ion cloud into the experimental setup.
More details on the ion production can be found in the methods section.
Before entering the \textsc{Alphatrap} magnet, the ion bunch passes a pulsed drift tube, where the kinetic energy is reduced to a few hundred $\SI{}{\eV} \times N_q$, which is necessary to capture the ions in the trap.
The cryogenic valve, shown in Fig.~\ref{fig:Trap}B, is opened briefly for the ion injection.
This way, the inflow of gas from the room-temperature beamline is blocked, achieving an ultra-high vacuum for long ion storage.
For this measurement campaign, a couple of hydrogenlike \textsuperscript{118}Sn\textsuperscript{49+} ions were loaded once.
One of these was stored for three months, which allowed to precisely measure the magnetic moment of the bound electron.
\\
The particles are trapped in our Penning-trap setup, which consists of a superconducting magnet with a $B$ field of roughly $\SI{4}{\tesla}$ for radial confinement.
This is overlapped with an electrostatic field which confines the ions in axial direction.
Once trapped, they are cooled via image currents to a temperature of $\SI{5.4(3)}{\kelvin}$.
In the magnetic field, the Zeemann effect splits the energy levels of the electron spin.
The energy difference is given as $h$ times the Larmor frequency $\nu_L=(g\,e\,B)/(4 \pi\,m_e) \approx\SI{107.6}{\giga\hertz}$, with $h$ the Planck constant, $g$ the $g$ factor, $e$ the electron charge and $m_e$ the electron mass.
Furthermore, the free-space cyclotron frequency $\nu_c=(q_\text{ion}\,B)/(2\pi\,m_\text{ion}) \approx\SI{25.7}{\mega\hertz}$ governs the motion of the stored ion, where $q_\text{ion}$ and $m_\text{ion}$ are its charge and mass, respectively.
Since both result from the magnetic field $B$, their relation allows access to the $g$ factor of the bound electron~\cite{brown_geonium_1986}:
\begin{equation}    
    g=2\,\frac{\nu_L}{\nu_c} \frac{q_\text{ion}}{e} \frac{m_e}{m_\text{ion}} = 2\,\Gamma_0\,N_\text{q} \frac{m_e}{m_\text{ion}}. 
    \label{eq:gfunc}
\end{equation}
The charge-ratio $N_q = q_\text{ion}/e$ is an integer number and the mass ratio is taken from other measurements~\cite{huang_ame_2021, tiesinga_codata_2021}.
This leaves the ratio $\Gamma_0=\nu_L/\nu_c$, which has to be experimentally determined in order to extract the $g$ factor.
In the presented measurement, the double-trap method is employed to determine $\Gamma_0$~\cite{haffner_double_2003}.
The 'trap stack' consists of two harmonic traps used for the measurement, and an additional section for ion capture and storage (see Fig.~\ref{fig:Trap}C).
In the precision trap, the three particle eigenmotions\begin{figure}[H]
\centering
    \includegraphics{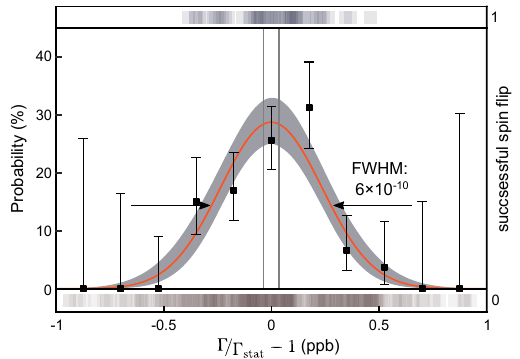}
    \caption{\textbf{Fig. 2: Measured spin-flip resonance of the bound electron in \textsuperscript{118}Sn\textsuperscript{49+}.} 
    The maximum-likelihood fit is shown as the orange line, together with its corresponding 1-sigma error-band (light grey).
    The scattered points are used to guide the eye, and represent a binned set of the data with \SI{68}{\percent} confidence levels (CL) given by a binomial fit.
    \SI{68}{\percent} CL for the resonance centre are shown as grey vertical lines.
    The square shadows above represent the successful spin-flips in the precision trap, the ones below show the unsuccessful attempts.
    }
    \label{fig:Resonance}
\end{figure} are determined. 
These are the modified cyclotron frequency $\nu_+\approx\SI{25}{\mega\hertz}$, the axial frequency $\nu_z\approx\SI{650}{\kilo\hertz}$ and the magnetron frequency $\nu_-\approx\SI{8}{\kilo\hertz}$.
These are in direct relation to the free-space cyclotron frequency via the invariance theorem $\nu_c^2= \nu_+^2+\nu_z^2+\nu_-^2$~\cite{brown_geonium_1986}.
The axial frequency is measured non-destructively by detection of image currents, induced by the moving particle next to the surrounding electrodes.
If the particle is in thermal equilibrium, the noise spectrum of the cryogenic detector shows a distinct 'dip' (see Fig.~\ref{fig:Trap}D).
The radial modes are detected by sideband coupling with the axial frequency, which enables the use of a single detector to measure all three frequencies.
To measure the spin orientation a second trap, called analysis trap, is used.
Its centre electrode is a ferromagnetic ring that produces a large quadratic coefficient of the magnetic field $B(z)=B_0+B_1z+B_2z^2+...$, with $B_2\approx\SI{45}{\kilo\tesla/\metre\squared}$.
Described as the continuous Stern-Gerlach effect~\cite{dehmelt_continuous_1986}, the electron spin interacts with this so-called magnetic bottle, resulting in a spin-dependent axial force.
Spin flips caused by irradiating microwaves in resonance with the Larmor-frequency $\nu_L$ can be detected in the analysis trap as a sudden change of the axial frequency as shown in Fig.~\ref{fig:Trap}E.
This magnetic field inhomogeneity is problematic for a precise $\Gamma$ measurement only in this trap, hence two traps optimised for their respective use allows significantly higher precision. 
\\
A measurement cycle starts in the analysis trap by determining the spin state.
Afterwards, the ion is adiabatically transported into the precision trap, where the particle eigenmotions are measured.
During the measurement of $\nu_+$, we irradiate a microwave at a random offset to the expected Larmor frequency.
Then, the ion is brought back into the analysis trap to probe whether the microwave injected in the precision trap has changed the spin orientation.
By repeating this at different offsets around the expected $\nu_L$, one gets a spin-flip probability as a function of the frequency ratio $\nu_L/\nu_c$.
More details are given in the Methods, and the measurement scheme is shown in Extended Data Fig. 1.
To determine the resonance parameters and their uncertainties a maximum-likelihood analysis is performed.
Multiple resonances have been recorded.
Most of these were performed with different settings and are used to check systematic effects like the relativistic correction.
For the extraction of $\Gamma_0$ only one is
\begin{table}[H]
    \centering
    \setlength\extrarowheight{1pt}
    \footnotesize
    \begin{tabularx}{\columnwidth}{l r r}
        \textbf{Parameter} & \textbf{Relative shift (ppt)}  & \textbf{Uncertainty (ppt)} \\ \hline\hline
        $\Gamma_0 = \nu_L/\nu_c$ error budget: & &\\ 
        \hspace{0.5cm}$\nu_-$ measurement & -\phantom{.0}\ & \SI{3.8}{}\\
        \hspace{0.5cm}Relativistic shift \cite{ketter_classical_2014} & \SI{23.7}{} & \SI{4.8}{}\\
        \hspace{0.5cm}Image-charge shift \cite{schuh_image_2019} & \SI{150}{}\phantom{.0}\ & \SI{7.5}{} \\ 
        \hspace{0.5cm}$\nu_z$ line shape & -\phantom{.0}\ & \SI{20}{}\phantom{.0}\\ 
        \hspace{0.5cm}statistical uncertainty & -\phantom{.0}\ & \SI{38}{}\phantom{.0}\\ \hline
        $g$-factor error budget: & &\\ 
        \hspace{0.5cm}Total $\Gamma_0$ uncertainty & &\SI{44}{}\phantom{.0}\\ 
        \hspace{0.5cm}Electron mass \cite{sturm_high-precision_2014, tiesinga_codata_2021} &  & \SI{29}{}\phantom{.0} \\ 
        \hspace{0.5cm}\textsuperscript{118}Sn\textsuperscript{49+} mass (this work)&  & \SI{475}{}\phantom{.0}\\   \hline \hline 
    \end{tabularx}
    \caption{\textbf{Table 1: Error Budget}
        The error budget of $\Gamma_0$ and $g$ is shown.
        Further contributions are smaller than \SI{1}{}~ppt, allowing to ignore them safely.
        More details can be found in the text and in the Methods section.
    }
    \label{tab:ExpData}
\end{table} used, as it is the most precise with small motional radii and weak microwave power.
\\
The scan consists of roughly 400 data points, 54 of these have been successful spin flips. 
The binned data, and the fit is shown in Fig.~\ref{fig:Resonance}.
It is not saturated, i.e. the maximum is well below \SI{50}{\percent}.
Therefore, the resonance shape is mostly determined by magnetic-field jitter, and not by power broadening.
We use a Gaussian fit function to analyse the resonance.
From this, $\Gamma_\text{stat}$ is extracted with a value of $\Gamma_\text{stat} = \SI{4189.05824237}{}(16)$.
The resulting ratio $\nu_L/\nu_c$ is corrected for systematic effects, arising from different sources as shown in the error budget in Tab. 1.
Further details are explained in the Methods.
The corrected $\Gamma_0$ amounts to:
\begin{equation}
    \Gamma_0 = \SI{4189.058241643}{}(160)_{\mathrm{stat}}(93)_{\mathrm{sys}}.
\end{equation}
The parentheses represent the statistical and systematic uncertainty, respectively.
Since the $g$ factor is also dependent on the mass of the highly charged ion, we additionally performed a cyclotron-frequency-ratio measurement to confirm the atomic mass evaluation (AME) value~\cite{huang_ame_2021}.
This yields a result of $m(\textsuperscript{118}\text{Sn}\textsuperscript{49+}) = \SI{117.874869069(56)}{}\text{u}$, improving the value obtained from the AME (corrected for the missing electrons and their binding energies) by roughly a factor of 10. 
\\
We also calculate the electron binding energies of neutral tin to extract the neutral tin mass to similar accuracy.
Details on the calculation and the mass measurement can be found in the Methods.
Using Eq. \eqref{eq:gfunc} we infer the $g$ factor to be:
\begin{equation}
    g_\text{exp} = \SI{1.910562058962}{(73)_{\mathrm{stat}}(42)_{\mathrm{sys}}}(910)_{\text{ext}}.
\end{equation}
All uncertainties are 1-sigma confidence levels. 
\begin{figure}[H]
\centering
    \includegraphics{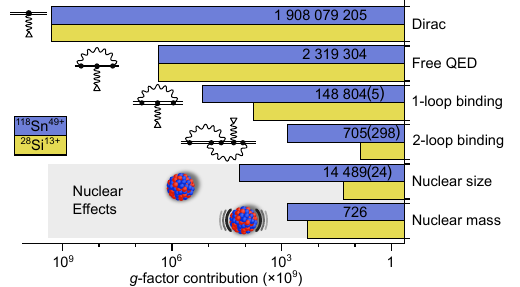}
        \caption{\textbf{Fig. 3: \textit g-factor contribution comparison of \textsuperscript{28}Si\textsuperscript{13+} versus \textsuperscript{118}Sn\textsuperscript{49+}.}
        Because of the significantly higher $Z\alpha$, the binding corrections to the $g$ factor increase strongly from Si ($Z=14$) to Sn ($Z=50$).
        In the values displayed without an uncertainty figure, all digits are significant.
    }
    \label{fig:SnSi}
\end{figure} 
\begin{figure}[H]  
\centering
    \includegraphics{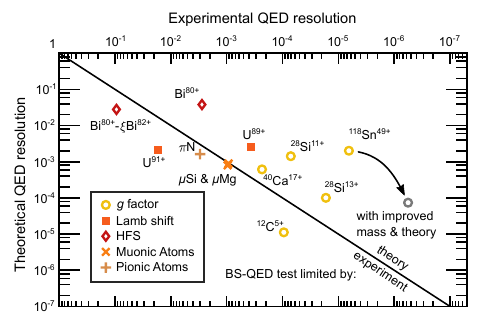}
    \caption{
        \textbf{Fig. 4: Bound-state QED tests in high electric fields.}
            The uncertainty of experiment and theory relative to the bound-state QED contribution for certain highlight measurements are shown~\cite{ullmann_high_2017, beiersdorfer_measurement_2005, gumberidze_quantum_2005, trassinelli_measurement_2016, beltrami_new_1986, kohler_isotope_2016, wagner_g_2013, sturm_g-factor_2013, sturm_high-precision_2014}.
            The separate measurements are summarised in the Extended Data Table 4.
            The muonic and pionic points are Lamb-shift measurements of the bound-muon/pion energy levels in an excited transition.
            Values above the diagonal line would profit from an improved theoretical calculation.
            In those below the line, the experimental error is the dominating uncertainty.
            The added limit with improved experiment and theory assumes a mass measurement that improves the $g$ factor to the $\Gamma_0$ uncertainty, as well as an improvement of the theoretical value to the limit imposed by the finite nuclear size uncertainty.
        }
    \label{fig:QEDtest}
\end{figure}
The brackets are respectively the statistical and the systematic uncertainty followed by the uncertainty of the external parameters, dominated by the atomic mass of tin-118.
Although the $\Gamma_0$ uncertainty is \SI{4.4e-11}{}, the remaining mass uncertainty of the \textsuperscript{118}Sn\textsuperscript{49+} ion limits the $g$ factor to a relative uncertainty of \SI{4.8e-10}{}.
\\
The theoretical description of the free-electron $g$ factor is well established~\cite{tiesinga_codata_2021}.

The dominant correction due to the binding Coulomb potential of the nucleus is described by the
Dirac value~\cite{breit_magnetic_1928}, $g_{\mathrm{D}} - 2 = \nicefrac{4}{3} \left( \sqrt{1- (Z\alpha)^2} -1 \right)$.
Apart from that, binding corrections of QED Feynman diagrams with closed loops need to be taken into account.
The non-relativistic QED approach which treats the interaction between electron and nucleus perturbatively (\cite{Pachucki2004oneloop}) cannot be expected to give good results for $Z=50$ because the expansion parameter of this perturbation series, $Z\alpha$, is too large.
Non-perturbative calculations for one-loop diagrams are well established~\cite{yerokhin_evaluation_2004, beier_g_j_2000}, while the calculations for two-loop diagrams are only partially done~\cite{yerokhin_two-loop_2013,sikora_theory_2020,debierre_two-loop_2021}.\\
The theory of the bound-electron $g$ factor has been previously tested in lighter ions, with $^{28}$Si$^{13+}$ being the heaviest hydrogenlike ion for which the $g$ factor has been measured~\cite{sturm_g_2011,sturm_g-factor_2013}.
In these previous measurements, one-loop binding corrections, namely the self-energy, the magnetic loop vacuum polarization and the Uehling part of the electric loop vacuum polarization corrections have been tested.
In this measurement of $^{118}$Sn$^{49+}$, for the first time in a $g$-factor measurement, the Wichmann-Kroll part of the vacuum polarization correction is larger than the total theoretical and experimental uncertainties.
Binding corrections to two-loop Feynman diagrams up to $\mathcal{O} \left(  (Z\alpha)^4 \right)$ were tested in previous measurements of  $^{28}$Si$^{13+}$~\cite{sturm_g_2011,sturm_g-factor_2013}.
Two-loop binding corrections of $\mathcal{O} \left(  (Z\alpha)^5 \right)$ which were calculated only after the $^{28}$Si$^{13+}$ measurements~\cite{czarnecki_two-loop_2018,czarnecki_logarithmically_2020}, turn out to be smaller than our estimated uncertainty due to uncalculated higher-order binding corrections.
In Fig.~\ref{fig:SnSi} different theoretical contributions to the bound-electron $g$ factor of $^{118}$Sn$^{49+}$ are presented and compared to the $^{28}$Si$^{13+}$ $g$ factor.
An extensive table, summarising the different contributions, is given in the Extended Data Tab. 3.
\\
In total, we find a theoretical \textsuperscript{118}Sn\textsuperscript{49+} $g$ factor of
\begin{equation}
g_\text{theo} = \SI{1.910561821(299)}{},
\end{equation}
in agreement with the experimental value, although with a much larger uncertainty, which is dominated by uncalculated higher-order binding corrections $\mathcal{O} \left( (Z\alpha)^6 \right)$ to two-loop Feynman diagrams.
Large-scale all-order calculations of these diagrams, which have the potential to greatly reduce the theoretical uncertainty, have been started in recent years~\cite{sikora_theory_2020,debierre_two-loop_2021}. 
\\Fig.~\ref{fig:QEDtest} shows the experimental- against the theoretical uncertainty for different tests of bound-state QED in systems with high electromagnetic fields.
Thus far, bound-state QED in heavy highly charged ions has been mostly tested by Lamb-shift measurements, the highest precisions were achieved in lithiumlike systems~\cite{beiersdorfer_measurement_2005,brandau_precise_2003,beiersdorfer_measurement_1998,beiersdorfer_structure_1995}.
With the tin measurement the underlying bound-state QED is tested to about \SI{0.20}{\percent}.
The total QED contribution, also including the zeroth order in the $Z\alpha$ expansion, is tested to about \SI{0.012}{\percent}.
As the test is purely limited by the estimated uncertainty of the uncalculated higher-order 2-loop terms, which is an order of magnitude larger than the uncertainty of effects from e.g. the finite nuclear size, the completion of ongoing calculations can potentially improve the QED test significantly. 
Additionally, an improved measurement of the atomic mass of the tin isotope could be achieved with higher precision by dedicated experiments (as shown in e.g.~\cite{kromer_high-precision_2022, schussler_detection_2020}), hence the experimental $g$-factor uncertainty can be reduced to that of $\Gamma_0$.
\\
In conclusion, the $g$-factor measurement of hydrogenlike \textsuperscript{118}Sn\textsuperscript{49+} paves the way for more sensitive tests of theoretical concepts and fundamental constants via Penning-trap precision $g$-factor measurements of highly charged ions.
It is a key step into the regime of strong fields previously uncharted for this kind of test.
By combining the production capabilities of the Heidelberg-EBIT with the high-precision Penning-trap setup \textsc{Alphatrap}, we demonstrated the suitability for numerous future $g$-factor measurements with heavy highly charged ions~\cite{shabaev_g-factor_2006, shabaev_ground-state_2022}.
Furthermore it marks the first steps towards hyperfine spectroscopy in a heavy highly charged ion with unprecedented precision, which could be done using a method similar to the one demonstrated in the laser spectroscopy of \textsuperscript{40}Ar\textsuperscript{13+}~\cite{egl_application_2019}.
Additionally, it is possible to measure different charge states and employ a weighted difference method in order to cancel finite size effects.
This, together with an improved theory, would allow more stringent QED tests, or possibly a determination of the fine-structure constant $\alpha$~\cite{shabaev_g-factor_2006,volotka_nuclear_2014,yerokhin_g_2016}.\newline \newline
We acknowledge help from Michael Rosner and Nils Rehbehn regarding the operation of the Heidelberg EBIT.
We also thank Natalia Oreshkina for discussion of the paper and its content.
This work was supported by the Max Planck Society (MPG), the 
International Max Planck Research School for Quantum Dynamics in Physics, Chemistry and 
Biology (IMPRS-QD), the German Research Foundation (DFG) Collaborative Research Centre 
SFB 1225 (ISOQUANT) and the Max Planck PTB RIKEN Center for Time, Constants, and 
Fundamental Symmetries. This project has received funding from the European Research 
Council (ERC) under the European Union’s Horizon 2020 research and innovation programme 
under grant agreement number 832848 FunI. 
This work comprises parts of the PhD thesis work of C.M.K. and J.M. to be submitted to Heidelberg University, Germany.
\end{multicols}



\begin{multicols}{2}    

\section{Methods}\label{Methods}
\subsection{Ion production}
As the K-shell electrons in \textsuperscript{118}Sn have a binding energy of around \SI{35}{\kilo\eV}, high energies are required for ionisation.
The ions produced in the EBIT are already rather slow ($7~\mathrm{keV} \times N_q$ with $N_q$ as the charge state) which simplifies the capture of ions after ejection.
In a pulsed drift tube, the kinetic energy is reduced to roughly $500~\mathrm{eV} \times N_q$, which is low enough that we can capture the ion bunch within the capture section of the trap.
A detailed description about the ion capture can be found in~\cite{sturm_alphatrap_2019}.
For the injection from the Heidelberg-EBIT, a lot of time was spent optimising the production efficiency as well as the ion transport through the beamline.
Currently the largest limit in the amount of ions that can be trapped in our experiment is the transport efficiency.
From the EBIT, a couple ten-thousand hydrogenlike Sn-118 are ejected.
In a single shot, due to poor transport efficiency in the beamline, at most a handful of particles can be trapped in our setup.
The particles get lost during transport through the beamline.
The phase-space distribution in the EBIT is rather large due to the high temperature in the plasma.
This results in a large spread of the ejected ions which makes it difficult to efficiently transport and trap the whole ion bunch.
Most ions are lost in the pulsed drift tube, where only a fraction of the bunch is decelerated.
Since this is a technical limitation, improvement in the transport efficiency would also be feasible in the future.
But since we perform the measurement mostly on single particles, we only have to trap a few particles, assuming the vacuum of the trap is good.
All together the required set-up time is about two weeks, with a following measurement time of multiple months.
For future measurements in even heavier hydrogenlike systems as e.g. hydrogenlike lead, the electron beam energy has to be increased further.
As of now, the Heidelberg-EBIT is limited to around \SI{65}{\kilo\V} of acceleration voltage, making the ionisation of higher binding energies impossible.
When this is overcome, either by improving the Heidelberg-EBIT or setting up an EBIT with enough electron-beam energy, the $g$-factor measurement is expected to be straightforward, as the measurement scheme for tin could also be used for e.g. an hydrogenlike lead or uranium measurement.
\subsection{Measurement scheme}
After capture, the next step is to perform the high-precision measurement and corresponding systematic checks.
Since this normally takes a few weeks, the vacuum in our trap has to be extremely good.
This is ensured by the cryogenic valve as shown in Fig.~1 which blocks the inflow of gas from the room-temperature beamline~\cite{sturm_alphatrap_2019}.
The valve and the cryogenic environment keep the vacuum below $10^{-16}~\mathrm{mbar}$, allowing ion storage of many weeks.\\
In the presented $\Gamma_0$ measurement, a method similar to~\cite{sturm_high-precision_2014} was used.
The measurement sequence is shown in Extended Data Fig.~1.
At first, the initial spin state is probed in the analysis trap (AT).
After adiabatic transport to the precision trap (PT) and a waiting time of two minutes to allow the voltages to settle, the modified-cyclotron frequency is determined by sideband coupling.
This is done by measuring and fitting of the noise spectrum that the detection circuit coupled to the ion produces~\cite{wineland_principles_1975}.
Afterwards, the axial frequency is measured with the detection circuit as well, a spectrum is shown in Fig.~1D.
This is followed by a 'pulse and amplify' (PnA) measurement sequence~\cite{sturm_phase-sensitive_2011}.
In PnA one determines the accumulated phase of the modified cyclotron motion after a fixed evolution time.
By initial excitation of the mode to a certain radius, a free evolving time for undisturbed phase accumulation followed by readout of the accumulated phase, one can track the phase of the particle during the evolution time, allowing to extract the frequency with high precision.
In this measurement, the set of evolution times consists of five reference phase measurements with 0.2 second evolution time, two unwrapping phases and two \SI{5.2}{\second} evolution times which are used for the determination of the magnetic field.
In the reference phases the magnetic field jitter is negligible on the phase stability, as this is not dominating yet.
For the long phases, the magnetic field stability is the dominating jitter/drift source.
The two unwrapping phases \SI{0.5}{\second} and \SI{2.2}{\second} are measured to allow a consistent phase unwrapping to the final \SI{5.2}{\second} measurement time.
All except the two \SI{5.2}{\second}-measurements are performed in random order.
The \SI{5.2}{\second} measurement is done twice at the end of the cycle.
The first is used to precisely determine the magnetic field, which is used to calculate the expected Larmor frequency for the microwave injection.
During the second \SI{5.2}{\second} measurement a microwave with a random frequency offset to the expected $\Gamma$ is irradiated.
With this, we measure the magnetic field one more time, while trying to perform a spin flip with the microwave.
For the final determination of the cyclotron frequency, we only use the five reference phases, and the last \SI{5.2}{\second} phase during the microwave excitation.
All others are used for unwrapping and to test certain systematics.
After the PnA cycle, the axial frequency is measured a second time.
The two axial frequency measurements are used to identify systematic shifts in the measurement, although mostly the second is used for the calculation of the free-space cyclotron frequency, as it is right after the microwave injection.
Afterwards, the ion is brought back to the AT, where we test if the PT microwave injection 'flipped' the spin.
The measurement scheme is shown in the Extended Data Fig.~1.
\subsection{Resonance analysis}
The resulting resonance was analysed with a maximum-likelihood fit.
Of the 387 spin-flip tries in the PT 54 have been successful.
The PT is at a position where the second order magnetic field inhomogeneity is smaller than \SI{10}{\milli\tesla/\metre\squared}, the influence on the field inhomogeneity can be safely neglected as the resulting error is below \SI{1}{ppt}~\cite{brown_geonium_1985}.
Odd-order inhomogeneity effects lead by the first order $B_1 \approx \SI{2.64(3)}{\milli\tesla/\metre}$ are even further suppressed since the odd orders are cancelled in a harmonic trap.
The remaining influences in the lineshape are given by magnetic field jitter, and by the power of the microwave injection.

Since the microwave injection time (\SI{5}{\sec}) is longer than the time the spin stays coherent to the (weak) drive, the spin-flip probability can be at most \SI{50}{\percent}.
If high microwave powers are used, the chance to drive a spin flip increases accordingly and broadens the resonance.
A microwave power dominated resonance would follow a Lorenzian line shape.
Vice versa, if the power is small enough, the magnetic field jitter dominates, the spin-flip probability drops below \SI{50}{\percent}, and the line shape changes from a Lorenzian to a Gaussian distribution.
For the described resonance a Gaussian function is used:
\begin{equation}
    P(\Gamma) = A e^{-\frac{\left(\Gamma-\Gamma_\text{stat}\right)^2}{2\sigma^2}},
\end{equation}
with $A$ being the amplitude, $\Gamma$ as the irradiated Larmor frequency divided by the measured cyclotron frequency, $\Gamma_\text{stat}$ as the centre of the resonance, and $\sigma$ the standard deviation of the normal distributed data.
Using a maximum-likelihood fitting method, which uses the unbinned dataset of $\Gamma$ ratios, we extract the parameters of the resonance.
It has an amplitude of $A = \SI{29(4)}{\percent}$, and a FWHM of \SI{5.6(3)e-10}{}.
Due to the low statistics, a Lorenzian fit results in a rather similar likelihood, but since the resonance width is consistent with the expected jitter based on the PnA phase stability, this is unlikely to be the case.
Further, the Lorenzian fit would give a smaller uncertainty on the resonance centre.
Using the Gaussian shape is thus the conservative approach in the extraction of $\Gamma_\text{stat}$ and its uncertainty.
In Extended Data Fig.~2 the likelihood surfaces of the maximum-likelihood fit are shown.
They show the maximum-likelihood planes in the three-dimensional parameter space spanned by the free parameters $\Gamma_\mathrm{stat}$, $\sigma$ and $A$.

\subsection{Systematic effects}
Of all the systematic effects, three dominate the uncertainty/correction.
The largest shift is the image-charge shift.
It is calculated according to Ref.~\cite{schuh_image_2019}, and amounts to a relative shift of \SI{1.50(8)e-10}{}.
Here, the error is \SI{5}{\percent} of the total value, following Ref.~\cite{schuh_image_2019}.
\\
The systematic effect with the largest uncertainty on $\Gamma_0$ arises from the axial frequency fit.
The axial resonator causes a frequency pushing depending on the relative frequency difference of particle and resonator.
This is accounted for in the fit function, but since the frequency of the resonator cannot be measured more accurately than a couple of Hertz, the resulting uncertainty in the lineshape of the axial dip can not be neglected.
With a resonator-frequency uncertainty of around \SI{2}{\hertz}, the fitted axial frequency changes by about \SI{20}{\milli\hertz}.
This translates to a \SI{2.0e-11}{} relative uncertainty on $\Gamma_0$.
\\
The relativistic correction due to the mass increase on a cyclotron radius of \SI{12.8(13)}{\micro\m} is \SI{2.4(5)e-11}{} (Ref.~\cite{ketter_first-order_2014}).
Here we assume an error of \SI{10}{\percent} on the cyclotron radius during PnA.
This was also confirmed by measuring a resonance with 161 cycles resulting in 18 spin flips at a significantly higher cyclotron radius (\SI{102(10)}{\micro\m}).
Due to the stronger mass increase, this is shifted by roughly \SI{1.6}{ppb}, and is in agreement with the expected shift (see Extended Data Fig.~3).
This resonance additionally tests for hypothetical systematic effects from the unwrapping, as in this dataset the longest PnA evolution time is set to \SI{10.2}{\second}, nearly twice as long compared to the main resonance.
\\
Due to the exceptionally harmonic trapping field in our PT and a good magnetic homogeneity, the residual shift of the result from higher-order coefficients in the electromagnetic potential can be limited to less than 1 part-per-trillion.
\\
Another uncertainty arises from the less frequent/precise measurement of the magnetron frequency.
This is only measured every couple 10 cycles, as its influence on the free-space cyclotron frequency is small.
Conservatively we assume a maximum error of \SI{0.3}{\hertz}, which results in a relative uncertainty of \SI{3.8e-12}{}.
An extensive table of the  different systematic effects is given in the Extended Data Table~\ref{tab:ExpandedExpData}.
\subsection{Mass measurement}
In order to exclude the Atomic Mass Evaluation (AME) value of the neutral \textsuperscript{118}Sn isotope as an error source to the $g$ factor, we performed a direct mass measurement.
We trapped a hydrogenlike \textsuperscript{118}Sn\textsuperscript{49+} ion along with a hydrogenlike \textsuperscript{12}C\textsuperscript{5+} ion.
Both have a relatively similar $q/m$ value, which results in suppressed systematic effects in the measurement.
The axial frequencies are separated by roughly \SI{820}{\hertz} for the same electric potential in the PT.
In order to determine the mass, we measure the cyclotron frequency ratio (CFR), which allows to extract the mass ratio without precise knowledge of the magnetic field:
\begin{multline}
    \frac{v_c(^{12}\text{C}^{5+})}{v_c(^{118}\text{Sn}^{49+})} = \frac{q(^{12}\text{C}^{5+})}{q(^{118}\text{Sn}^{49+})} \frac{m(^{118}\text{Sn}^{49+})}{m(^{12}\text{C}^{5+})} = \\ \frac{5}{49} \frac{m(^{118}\text{Sn}^{49+})}{m(^{12}\text{C}^{5+})}.
\end{multline}
To determine the mass, we employed two measurement methods with partially different systematic effects to cross check the result.
A sketch of the measurement schemes can be found in the Extended Data Fig.~4.
In both we measure the ions in an interleaved manner, transporting each particle subsequently into the PT, while storing the other in a neighbouring section.
In the main method we employ the 'pulse and amplify' (PnA) scheme in order to determine the modified cyclotron frequency without any lineshape uncertainty.
Atypically, we perform this measurement with one particle on resonance, and the other \SI{820}{\hertz} apart.
That way, we can apply the same voltages to the electrodes, and effects due to a shifted potential/ion position are avoided.
For this, we put the carbon ion in resonance, as due to lower charge, this couples less to the detector system.
It would be virtually impossible to detect its image currents when detuned by that amount.
But due to the stronger coupling in hydrogenlike tin, this is still detectable even \SI{820}{\hertz} detuned, allowing to measure the motional frequencies.
\\
In the second method, we change the electrostatic potential to tune each axial frequency into resonance with the detector.
We use sideband coupling, also known as a 'double dip' measurement, to measure the modified cyclotron frequencies of each particle.
In order to move each particle on resonance, the axial potential is changed by roughly \SI{150}{\milli\volt} out of the initial $V \approx \SI{-59}{\volt}$.
As possible patch potentials on the electrodes are not necessarily symmetric, the trap centre might shift slightly due to the different voltages.
Combined with a magnetic field gradient $B_1 = \SI{2.64(3)}{\milli\tesla/\metre}$ (Ref.~\cite{sturm_alphatrap_2019}), the measured frequency ratio could be systematically shifted due to this possible misalignment.
\\
Another possible systematic shift arises from the different lineshapes of the ions on the detector.
The dip width of hydrogenlike tin and hydrogenlike carbon differs by about a factor of 10.
Thus, effects due to an erroneous input parameter to the fitting model of the 'double dip' could result in a systematic effect, as each ion would be affected differently.
In the PnA method, the modified cyclotron frequency is determined independent of the detector line shape, therefore removing the connected systematic uncertainties.
Due to the better understanding of the systematic effects in the PnA method, we use the 'double dip' measurement solely as a cross-check of the measured mass.
When considering all systematic effects and uncertainties, this agrees with the PnA measurement.

Here is to note, that only five frequency ratios have been collected in the phase-sensitive measurement, therefore we conservatively use the standard deviation as our statistical uncertainty, and not the standard deviation of the mean.

Correcting the measurement for relativistic effects and the image charge shift, we get a final value for the mass ratio:
\begin{equation}
    \begin{aligned}
        \frac{m\left(^{118}\text{Sn}^{49+}\right)}{m\left(^{12}\text{C}^{5+}\right)} &= \SI{9.8251510645}{}(39)_\mathrm{stat}(27)_\mathrm{sys}.
    \end{aligned}
\end{equation}
The two brackets represent the statistical and systematic uncertainty respectively.
The mass of the \textsuperscript{12}\text{C}\textsuperscript{5+} ion can be expressed in relation to the neutral carbon atom as unit of mass after accounting for the missing electrons and binding energies.
This gives a mass value for the \textsuperscript{12}\text{C}\textsuperscript{5+} ion of \SI{11.99725768029217}{}(43)(8)u (Ref.~\cite{tiesinga_codata_2021, kramida_nist_2021}).
The brackets are the uncertainty of the binding energies and the electron mass uncertainty respectively.
As the total uncertainty is below 0.01 parts-per-trillion, its influence on the resulting hydrogenlike tin mass can be safely neglected.
From this we can infer a value for the atomic mass of the hydrogenlike tin-118 ion:
\begin{equation}
    \begin{aligned}
        m\left(^{118}\text{Sn}^{49+}\right) &= \SI{117.874869069}{}(47)_\mathrm{stat}(32)_\mathrm{sys} \text{u}.
    \end{aligned}
\end{equation}
The first bracket is the statistical uncertainty, the second the systematic uncertainty, which is dominated by the relativistic effect.

\subsection{Binding energies and neutral mass}
The AME2020 value for the mass of neutral $^{118}$Sn is $\SI{117.90160663(54)}{}$ \,\text{u}, which has an uncertainty of 466~eV/c$^2$ (u=$\SI{9.3149410242(28)e8}{\eV/c^2}$). With the measured mass of $^{118}$Sn$^{49+}$, one can improve the accuracy of the mass of the neutral atom via
\begin{eqnarray}
m\left({}^{118}\text{Sn}\right)&=&m\left({}^{118}\text{Sn}^{49+}\right)+49m_e-\Delta E/c^2.
\end{eqnarray} 
Here, $m_e=\SI{0.000548579909070(16)}{}$ \,\text{u} is the electron rest mass~\cite{tiesinga_codata_2021}. $\Delta E$ is the energy required to ionize the 49 electrons from a neutral Sn atom, and is theoretically calculated to be 132746(5)~eV. The final result is $m\left({}^{118}\text{Sn}\right)=\SI{117.901606974(56)}{}(5) \,\text{u}$, which is a factor of 9.5 more accurate than the previous best value.

The value $\Delta E$ is derived from the electron-binding-energy difference between neutral Sn and hydrogen-like Sn$^{49+}$.
As the electron binding energy of Sn$^{49+}$ is known as $35192.501(11)$~eV~\cite{yerokhin_lamb_2015}, we only need to calculate the electron binding energy of neutral tin.
This is performed with an \textit{ab-initio} fully relativistic multiconfiguration Dirac--Hartree--Fock (MCDHF) together with a relativistic configuration interaction (RCI) method~\cite{grant_relativistic_1970,desclaux_relativistic_1971,GRASP2018} implemented in the GRASP2018 code~\cite{GRASP2018} . 
However, since the binding energy of the outermost 4~electrons of Sn has been experimentally determined to be 93.22(4)~eV~\cite{kramida_nist_2021}, the binding energy of the ground state of Pd-like Sn$^{4+}$ ([Kr]$4d^{10}$~${}^1S_0$) is calculated instead. 
We note that the ionisation potential (IP) for the fifth electron is also known experimentally.
However, the ground state of Sn$^{5+}$ is an open-shell conﬁguration with a total angular momentum of 5/2. Thus, it requires a much larger basis set and computational power to achieve the same accuracy as for the closed-shell ion Sn$^{4+}$.

Within the MCDHF scheme, the many-electron atomic state function (ASF) is constructed as a linear combination of configuration state functions (CSFs) with common total angular momentum ($J$), magnetic ($M$), and parity ($P$) quantum \mbox{numbers:} $\lvert \Gamma P J M\rangle = \sum_{k} c_k \lvert\gamma_k P J M\rangle $.
The CSFs $\lvert\gamma_k P J M\rangle $ are given as $jj$-coupled Slater determinants of one-electron orbitals, and $\gamma_k$ summarizes all the parameters needed to fully define the CSF, i.e., the orbital occupation and coupling of single-electron angular momenta.
$\Gamma$ collectively denotes \mbox{all} the $\gamma_k$ included in the representation of the ASF.
The mixing coefficients $c_k$ and the radial orbital wave functions are obtained by solving the MCDHF equations self-consistently~\cite{grant_relativistic_1970,desclaux_relativistic_1971} including the Dirac--Coulomb Hamiltonian.
After that, the RCI method is employed to calculate the contributions from mass shift, transverse photon interactions and QED effects. 

We start with a Dirac--Hartree--Fock (DHF) calculation where only the ground state configuration is considered.
This gives a binding energy of 167973.14~eV, with a -4.42-eV correction from the ﬁnite nuclear size effect.
Further calculation with RCI adds contributions of -0.57~eV from mass shift, -120.34~eV from frequency-independent transverse photon interactions (or Breit interactions), 1.16~eV from frequency-dependent transverse photon interactions and -79.08~eV from QED terms. 
To derive the electron correlation energy, the size of the CSF basis set is gradually expanded via single and double (SD) excitation of electrons from the ground state configuration to high-lying virtual orbitals.
This allows us to monitor the convergence of the correlation energy by adding and optimizing virtual orbitals layer by layer up to $n=10$ ($n$ is the principal quantum number). 

Six terms may lead the error bar: the uncertainty in the nuclear parameters, the finite basis set, the uncounted higher-order electron correlations, the insufficient basis functions, the inaccurate estimations for QED corrections and the uncalculated QED effect to the mass shift.
The uncertainty in the nuclear radius gives a 0.06-eV uncertainty to the corrections in the ﬁnite nuclear size effect.
To exclude the error caused by the finite basis set, we extrapolate our calculated value to $n=\infty$.
This results in a SD correlation energy of 67.26(23)~eV for the ground state of Sn$^{4+}$.
Since this analysis is performed under the RCI calculation, the contribution from Breit interaction is fully accounted for.
The frequency-dependent transverse photon interaction can not be included in the multiconﬁguration calculations of GRASP2018, thus its uncertainty will be accounted for later together with other untreated minor effects when deriving the systematic errors.
The relativistic mass-shift operator implemented in the GRASP2018 code is accurate to the order of $(m_e/M)(\alpha Z)^4m_ec^2$ ($M$ is the mass of the nucleus).
Therefore, it bears an uncertainty of 0.02~eV for the mass-shift correction. 

In the GRASP2018 code, the QED corrections are estimated via a screened-hydrogenic approximation~\cite{GRASP2018}. 
This correction is dominated by inner shell electrons.
With an absolute value of 79.08~eV for Sn$^{4+}$, one already has a QED correction of 76.64~eV for Be-like Sn$^{46+}$. 
Fortunately, the QED effect for a nearby element, Be-like Xe$^{50+}$, has been known to a sub-eV accuracy via \textit{ab-initio} QED calculations~\cite{malyshev_qed_2014,malyshev_ab_2021}.
This allows us to infer the accuracy of our QED calculations: with a calculated value of 99.03~eV for Xe$^{50+}$, it is 0.85-eV larger than its \textit{ab initio} result.
Assuming the same relative deviation, we derive a QED correction of 78.40(68)~eV for Sn$^{4+}$. 
However, for such an ion, the many-electron QED effects are difﬁcult to evaluate accurately.
In the following, we will effectively include these contributions into the systematic errors.

The systematic errors caused by the uncalculated effects can be estimated from the ionization potentials (IPs) of the outermost electrons of Sn.
For example, the IP of the $5s$ electron in Sn$^{3+}$ ([Kr]$4d^{10}5s$~${}^2S_{1/2}$) is determined to be 40.74(4)~eV by experiment~\cite{kramida_nist_2021}.
With the calculated binding energy of Sn$^{3+}$ under a similar SD excitation scheme, we derive an IP of 40.07~eV which is 0.67~eV smaller than the experimental value.
This deviation shall be mainly originated from the high-order correlation effects, many-electron QED effects, insufﬁcient basis functions, and frequency-dependent transverse photon interactions.
Nevertheless, this deviation becomes smaller for highly charged ions~\cite{kromer_high-precision_2022}.
Therefore, one could conservatively assume that the corrections to the IPs decrease linearly to 0.10~eV for Cu-like Sn$^{21+}$ (since we have found that the deviation is already below 0.10~eV for the IP of Cu-like Kr$^{6+}$).
For the IPs of ions throughout Sn$^{21+}$ to Sn$^+$, we conservatively assume that they all have a correction of 0.10 eV.
In total, the contribution from all unaccounted for terms shall be within 0 and 9.73~eV. 
To cover this whole range, one can add a correction of 4.86~eV to the total binding energy of Sn$^{4+}$ and simultaneously assume a systematic error of 4.86~eV.

Finally, we arrive at a total binding energy of 167847(5)~eV for the ground state of Sn$^{4+}$. 
The different contributions and their uncertainties are summarised in the Extended Data Table~2.
This gives a binding energy difference of 132748(5)~eV between neutral Sn and hydrogen-like Sn$^{49+}$.
Combining the measured mass for Sn$^{49+}$, we obtain 
\begin{equation}
    \begin{aligned}
        m\left({}^{118}\text{Sn}\right)=\SI{117.901606974(56)}{}_{\mathrm{exp}}(5)_{\mathrm{theo}} \,\text{u}
    \end{aligned}
\end{equation}
for the neutral tin-118 atom.
The first bracket is the measurement uncertainty, while the second is the uncertainty of the electron binding energies.

\subsection{Theory of the bound-electron \textit g factor}

The leading $g$-factor contribution was first calculated in Ref.~\cite{breit_magnetic_1928}.
It is based on the approximation of an infinitely small and infinitely heavy nucleus.
Therefore, corrections due to the finite nuclear size and mass need to be taken into account.
Furthermore, QED corrections contribute to the total $g$ factor, just as in the case of the free electron.
However, QED corrections in case of the bound electron differ from the free-electron case.
In Fig. 4, we highlight especially QED binding corrections, i.e. the difference between QED corrections for bound and free electron. 
In the following, we discuss all relevant contributions.

\subsection*{Free-electron contributions}
Contributions to the free-electron $g$-factor were taken from Ref.~\cite{tiesinga_codata_2021}, namely
the one- to five-loop QED contributions.
Hadronic as well as electroweak contributions as given in~\cite{tiesinga_codata_2021} are too small to be relevant in this work.
\subsection*{Nuclear corrections}
The finite size (FS) correction to the $g$ factor as given in Extended Data Table 3 was calculated for the two-parameter Fermi distribution using formulas and tabulated parameters from Ref.~\cite{Yerokhin2013}.
The first uncertainty corresponds to the nuclear root-mean-square (rms) charge radius as given in~\cite{angeli_table_2013}.
Note that the uncertainty corresponding to the number of digits specified for relevant parameters in~\cite{Yerokhin2013} is much smaller than the radius uncertainty, $\SI{7e-10}{}\ll\SI{2e-8}{}$.
The second uncertainty of the FS correction from the table is a combination of the uncertainty due to the nuclear polarization, deformation and susceptibility, together with a conservative estimate of the nuclear model dependence.
The model dependence is the leading uncertainty, and expresses the difference of the FS corrections for the two-parameter Fermi and the homogeneously charged sphere model, again following Ref.~\cite{Yerokhin2013}.
A direct calculation of the FS correction for the sphere model using semi-analytic wave functions was consistent with the result from~\cite{Yerokhin2013}.\\
Calculating the FS correction analytically using formulas from Refs.~\cite{Zatorski12,Karshenboim2018}, we find a disagreement with numerical results corresponding to 3 times the nuclear radius uncertainty for both the sphere and Fermi models.
This suggests that additional higher-order contributions need to be determined for the analytic approach to be accurate at high $Z$ such as $Z=50$.\\
Finally, we also calculated the FS correction using the GRASP code~\cite{GRASP2018}.\\
In the absence of results for $^{118}$Sn$^{49+}$ in Refs.~\cite{Nefiodov02,cakir_improved_2020}, the nuclear polarization correction was estimated as zero following Ref.~\cite{Yerokhin2013}, with an uncertainty estimated as 50\% of the nuclear model uncertainty of the FS correction.
Our estimates for the additional uncertainty due to nuclear deformation~\cite{michel_nonperturbative_2019} and nuclear susceptibility~\cite{jentschura_mass_2006} corrections are negligible.\\
For recoil calculations, we used the mass value presented in this paper.
The leading recoil term of first order in the mass ratio was calculated to all orders in $Z\alpha$ using formulas and tabulated parameters in Ref.~\cite{Shabaev02}.
Recoil corrections of higher-order in $\frac{m}{M}$ were calculated to the leading order in $Z\alpha$ but exactly in the mass ratio~\cite{Close1971, Eides1997,Pachucki08}.
We find the result up to $\mathcal{O} \left( \left( \frac{m}{M} \right)^2 \right)$~\cite{Faustov1970,Grotch1971} to deviate by less than $1\times 10^{-14}$ from the all-order result.

Radiative recoil corrections were calculated using formulas from refs.~\cite{Grotch1971, Close1971,Faustov1970,beier_g_j_2000, Pachucki08}.
So far, recoil corrections to the $g$ factor have been derived only for the model of a point-like nucleus.

\subsection*{One-loop QED}
Binding corrections at the one-loop level have been calculated to all orders in $Z\alpha$.
For the one-loop self-energy correction, we used the result from Ref.~\cite{Yerokhin04} which is based on the model of a point-like nucleus.
We calculated the Uehling part of the electric loop vacuum polarization diagram for the model of a point-like nucleus both numerically, using the Uehling potential from Ref.~\cite{peskin_introduction_2018} and the bound electron wavefunction perturbed by a constant external magnetic field as derived in~\cite{ShabaevVirial}, as well as analytically using formulas from Ref.~\cite{Karshenboim2001}, with both results in excellent agreement.
Values for the Wichmann-Kroll electric loop vacuum polarization (VP) correction as well as the magnetic loop VP correction were taken from Ref.~\cite{beier_g_j_2000} and were calculated for extended nuclei in that work.

\subsection*{Combined QED-FS corrections}

Bound-electron QED corrections, when carried out for the model of an extended nucleus, give slightly different results compared to point-nucleus calculations.
Therefore, for an accurate theoretical description of the bound-electron $g$ factor, one has to take into account the correction to QED contributions due to the finite nuclear size (QED-FS corrections).
As mentioned in the previous section, the Wichmann-Kroll part of the electric loop vacuum polarization as well as the magnetic loop vacuum polarization corrections already include the QED-FS corrections.
We calculated the finite size corrections to the one-loop self-energy correction and the Uehling contribution to
the electric-loop vacuum polarization correction for the two-parameter Fermi distribution of the nuclear charge using formulas, and tabulated parameters for $Z = 50$, as given in Ref.~\cite{Yerokhin2013}.
An older calculation of the FS correction to the one-loop SE contribution based on the spherical shell model of the nucleus from Ref.~\cite{yerokhin_evaluation_2004} differs by $2.1 \times 10^{-9}$ from the result for the Fermi distribution.
Using semi-analytic wavefunctions for the homogeneous sphere model of the nucleus~\cite{Trouyet2019}, we calculated the FS correction to the Uehling part of the electric loop vacuum polarization correction~\cite{Karshenboim2005}.
Our result differs by $5 \times 10^{-10}$ from the result for the Fermi distribution.
    We therefore assign (nuclear model) uncertainties of \SI{2.1e-9}{} to the one-loop self-energy FS correction as well as \SI{5e-10}{} to the FS-correction to the Uehling part of the electric-loop vacuum polarization correction.

\subsection*{Two- and higher-loop QED diagrams}
We calculated binding corrections to two- to five-loop QED diagrams of order $(Z\alpha)^2$ following Ref.~\cite{czarnecki_anomalous_2000}.
See also refs.~\cite{Grotch1971,Close1971,Faustov1970} for earlier derivations of these binding corrections. (Results given in the lines “2-loop QED, $(Z\alpha)^2$”
and “$\geq$3-loop QED binding”)\\
Two-loop binding corrections of order $(Z\alpha)^4$ were derived in refs.~\cite{pachucki_complete_2005, czarnecki_two-loop_2018}.
All-order calculations in $Z\alpha$ were carried out in Ref.~\cite{yerokhin_two-loop_2013} for a subset of two-loop diagrams, namely those diagrams with at least one vacuum polarization loop.
However, magnetic loop vacuum polarization diagrams were not considered in that work.
In Ref.~\cite{yerokhin_two-loop_2013}, results are given explicitly for the contribution of orders $(Z\alpha)^5$ and higher.
We give this result in the line “S(VP)E, SEVP, VPVP”.
For the remaining two-loop Feynman diagrams, QED corrections of order $(Z\alpha)^5$ were calculated in Ref.~\cite{czarnecki_two-loop_2018}.
In Ref.~\cite{czarnecki_two-loop_2018}, a relative uncertainty of $13\%$ is mentioned, due to uncalculated Feynman diagrams contributing to order $(Z\alpha)^5$.
This corresponds to our uncertainty in the line $(Z\alpha)^5$ in Extended Data Table 3.
The uncertainty in the line $(Z\alpha)^{5+}$ S(VP)E, SEVP, VPVP corresponds to higher-order corrections of order $(Z\alpha)^{6+}$ of two-loop Feynman diagrams with one vacuum polarization magnetic loop, by interpolating between tabulated results from Ref.~\cite{debierre_two-loop_2021} for nearby $Z$.\\
The uncertainties due to uncalculated Feynman diagrams with two self-energy loops of order $(Z\alpha)^{6+}$ were estimated using the methods from Refs.~\cite{Pachucki05} and~\cite{czarnecki_two-loop_2018}, with the larger of the two estimates chosen as the uncertainty for these contributions in Extended Data Table 3.
Furthermore, uncalculated binding corrections of $\mathcal{O} \left( (Z\alpha)^{4+} \right)$ to Feynman diagrams with three and more loops were estimated by adapting the method from Ref.~\cite{Pachucki05} (uncertainty in the line “$\geq$3-loop QED binding”).\\
As can be seen from Extended Data Table 3, the total uncertainty of the theoretical $g$-factor value is dominated by uncalculated higher-order in $Z\alpha$ two-loop QED contributions.
Calculations to improve the accuracy of two-loop corrections are underway~\cite{sikora_theory_2020,debierre_two-loop_2021}. 
In order to improve the theoretical accuracy, an all-order (in $Z\alpha$) calculation of the two-loop QED correction is required.
Such calculations are underway~\cite{debierre_two-loop_2021, sikora_theory_2020}.
The most difficult part of the calculation is the two-loop self-energy, which is split into several parts according to the degree of their ultaviolet divergence, which are known as the loop-after-loop (LAL) correction, the F-, M-, and P-terms~\cite{Mallampalli1998}.\\
The F term is the part with overlapping ultraviolet divergences.
It can be represented by Feynman diagrams with only free electron propagators inside the self-energy loops and is evaluated in momentum space, thus avoiding any partial-wave expansion.\\
The M-term is the ultraviolet finite part of the two-loop self-energy correction with Coulomb Dirac propagators inside the self-energy loops.
Since the Coulomb Dirac propagator is best known in coordinate space, M-term calculations need to be carried out using a coordinate representation.
Typically, M-term contributions are a double infinite sum of partial waves over angular momentum quantum numbers which requires a very large number of partial waves to be calculated in practice.
The calculation of every partial wave requires a multidimensional integration to be carried out numerically.\\
The P-term is the part of the two-loop self-energy correction which contains Coulomb Dirac propagators in one part of the Feynman diagrams as well as an ultraviolet divergent subdiagram.
This requires P-term contributions to be calculated in a mixed coordinate-momentum representation, which involves the numerical Fourier transforms of the Dirac-Coulomb propagators over one of the radial arguments.
Details can be found in our earlier work~\cite{sikora_theory_2020}.

Results for the so-called loop after loop (LAL) and F-term contributions to two-loop self-energy corrections have been obtained, with the uncertainty of the LAL correction given in Ref.~\cite{sikora_theory_2020} for $Z=50$ being $6.5 \times 10^{-9}$, and the uncertainty of the F-term being orders of magnitude better.
The calculation of the remaining parts of the two-loop self-energy correction, i.e. the M- and P-terms, is ongoing.
A significant improvement of the total theoretical uncertainty compared to the value given in Extended Data Table 3 can be anticipated once these calculations are complete.
\subsection*{Muonic and hadronic vacuum polarization}
The result for the muonic vacuum polarization correction given in Extended Data Table 3 corresponds to the Uehling part of the electric loop contribution, calculated for the model of a point-like nucleus.
Comparing results for point-like nuclei with results for extended nuclei from Ref.~\cite{Belov2016}, we find good agreement between both for low $Z$.
For $Z=70$, the result for the extended nucleus is only about $50\%$ of the point-nucleus result.
For even higher $Z$, the extended nucleus result is significantly smaller than $50\%$ of the point-nucleus result.
For $Z=50$, we therefore expect the muonic vacuum polarization correction for an extended nucleus to be larger than $50\%$ of the point-nucleus result, and assign a $50\%$ uncertainty to the muonic vacuum polarization correction.
We also calculated the Uehling part of the muonic vacuum polarization correction for the sphere model of the nucleus~\cite{Trouyet2019,Moutet2020,Karshenboim2005}.
Our result of $-2.0 \times 10^{-9}$ is, within the specified uncertainty, in agreement with the point-nucleus result from Extended Data Table 3.\\
The hadronic vacuum polarization correction (\enquote{Hadronic Uehling}) was estimated following Ref.~\cite{Friar1999,Belov2016}, as 0.671 and 0.664 times the muonic vacuum polarization correction with both estimates being identical to all digits given.
We assigned the same uncertainty to the hadronic vacuum polarization correction which was used for the muonic vacuum polarization correction.
\end{multicols}

\subsection{Extended Data}
\begin{table}[H]
    \centering
    \setlength\extrarowheight{2pt}
    \begin{tabular}{l l l}
        \multicolumn{2}{l}{\textbf{Fit parameter}} & \textbf{value}  \\ \hline\hline
        & Amplitude & \SI{29(4)}{\percent}\\ \hline
        & Full-Width-Half-Maximum & \SI{5.6(3)e-10}{}\\ \hline
        & $\Gamma_\text{stat}$ & \SI{4 189.058 242 35(16)}{} \\ \hline 
        \multicolumn{2}{l}{\textbf{Systematic shifts}} & \textbf{relative shift}  \\ \hline\hline
        & Image charge shift & \SI{1.50(8)e-10}{}\\ \hline
        & Relativistic shift & \SI{2.37(48)e-11}{}\\ \hline 
        \multicolumn{2}{l}{\textbf{systematic uncertainties} (not shifts)} & \textbf{relative uncertainty}  \\ \hline\hline
        & \textsuperscript{118}Sn mass (this work) & \SI{4.8e-10}{}\\ \hline    
        & Electron mass \cite{tiesinga_codata_2021} & \SI{2.9e-11}{}\\ \hline
        & Axial dip line shape & \SI{2.0e-11}{}\\ \hline 
        & $\nu_-$ frequency uncertainty & \SI{3.8e-12}{}\\ \hline 
        & Electrostatic anharmonicity $C_4$, $C_6$, ... & \SI{<6e-14}{}\\ \hline 
        & $B_2$ line shape & \SI{<6e-14}{}\\ \hline 
    \end{tabular}
    \caption{\textbf{Extended Data Table 1: Experimental results} The fit parameters for the resonance are given, as well as the dominating systematic effects and the dominating uncertainties for the resulting $g$ factor}
    \label{tab:ExpandedExpData}
\end{table}
\begin{figure}[H] 
    \centering
    \includegraphics[width=\textwidth]{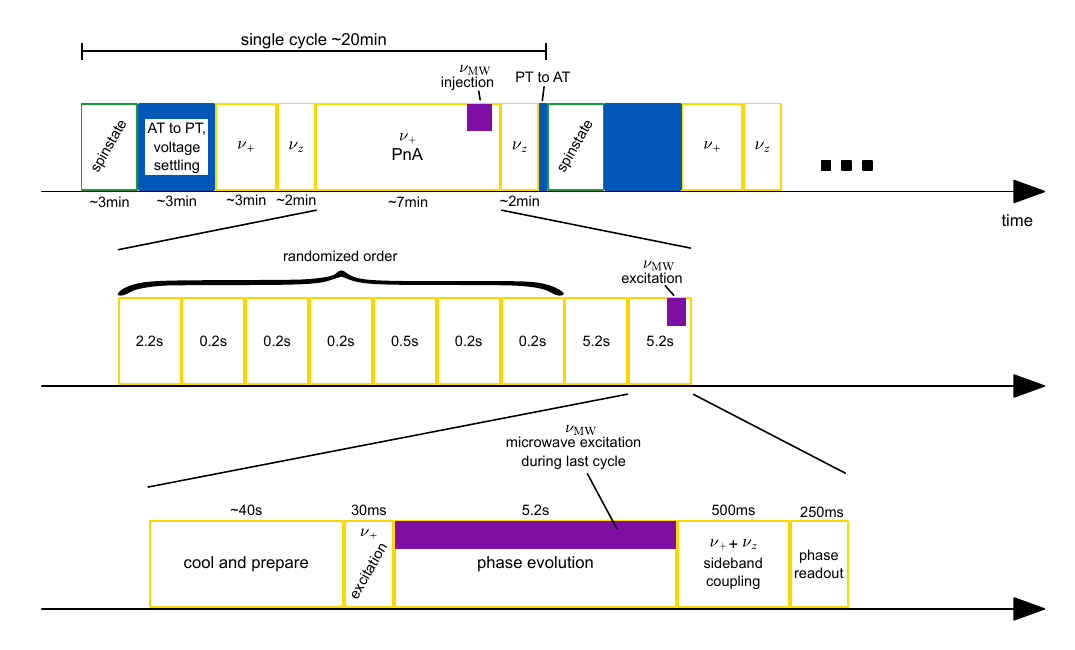}
    \caption{\textbf{Extended Data Fig. 1: $\mathbf{\Gamma}$ measurement scheme.}
        The ratio $\Gamma = \nu_L/\nu_c$ is measured with the shown scheme, $\nu_+$ is measured most precisely via the PnA method.
        $\nu_-$ is measured only every couple of cycles, as only moderate precision is required for a precise $\nu_c$ measurement.
        In the PnA method, different evolution times are used.
        The first few are randomised, while the last two are always with 5.2~s evolution times.
        The first precisely determines the current $\nu_c$, which is used to guess the $\nu_c$ for the microwave injection with higher accuracy.  
    }
    \label{fig:gscheme}
\end{figure}
\clearpage
\begin{figure}[H]  
    \centering
    \includegraphics{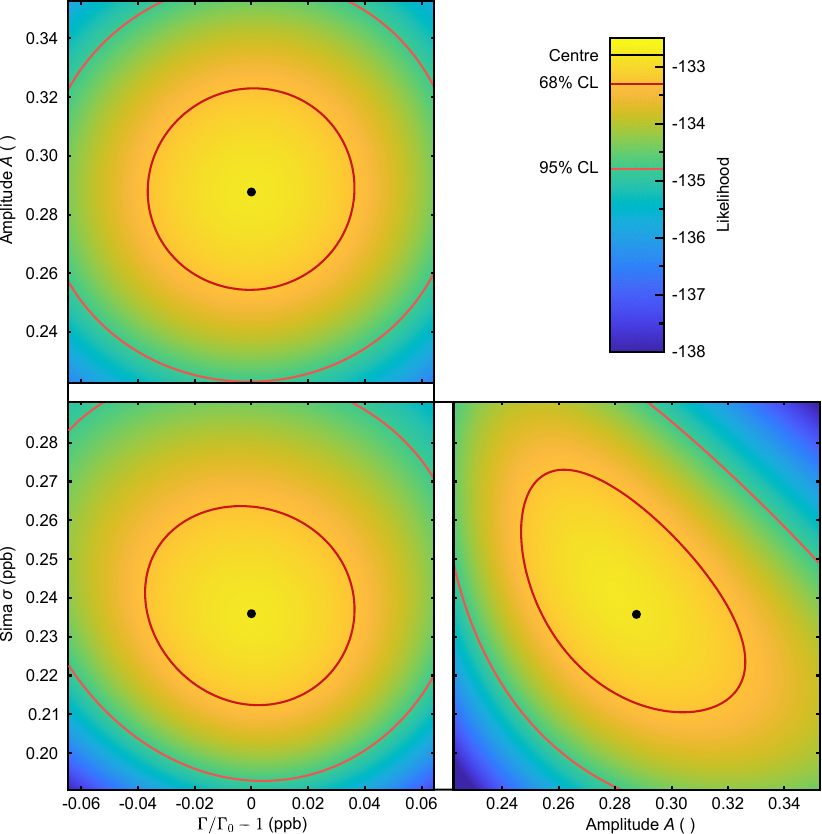}
    \caption{\textbf{Extended Data Fig. 2: Likelihood Surfaces}
        The likelihood planes through the optimal parameter-set (black circles) are shown.
        In the $\Gamma_0$ dimension, the surfaces show a circular shape, indicating little to no correlation with the other parameters.
        Overall, the fit converged correctly to the global maximum-likelihood position.
    }
    \label{fig:MLfunc}
\end{figure}
\begin{figure}[H]  
    \centering
    \includegraphics{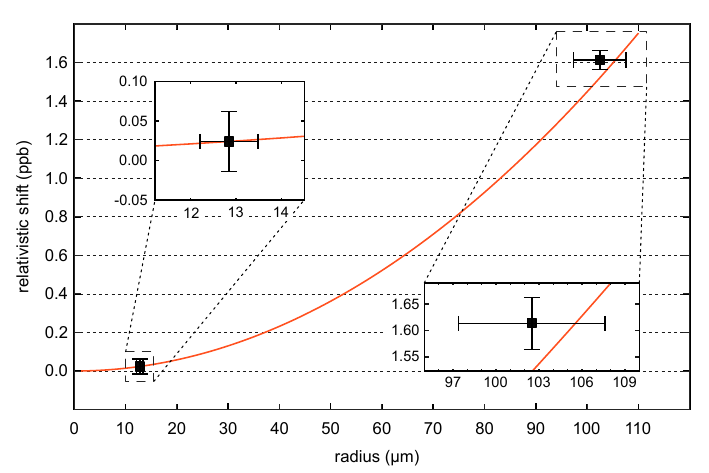}
    \caption{\textbf{Extended Data Fig. 3: Relativistic shift}
        Two resonances with different cyclotron radii have been measured.
        One with a cyclotron radius of \SI{12.8(13)}{\micro\m}, and the other with \SI{102(10)}{\micro\m}.
        Both resonances are in agreement with the total relativistic shift for a given radius (red line).
    }
    \label{fig:RMS}
\end{figure}
\begin{table}[H]
    \setlength{\tabcolsep}{6pt}
\begin{center}
\begin{tabular}{lrrrrrrrrr}
  \hline
 & DHF$_0$ & FNS &  Breit & $\omega$TP &  MS & QED & SDc & HO & total\\
 \hline
$E_{\text{Sn}^{4+}}$ & $167977.60$ & $-4.42$ &  $-120.34$ & $1.16$ & $-0.57$ &    $-78.40$ & $67.26$ & $4.86$ & $167847.15$ \\
$\delta E_{\text{Sn}^{4+}}$ & NaN & $0.06$ &  NaN & NaN & $0.02$ &    $0.68$ & $0.23$ & $4.86$ & $4.92$ \\ 
\end{tabular}
\caption{\textbf{Extended Data Table 2: Binding Energy} Different contributions to the binding energy of Sn$^{4+}$: DHF$_0$ being the DHF energy assuming a point-like nuclear charge, FNS the finite nuclear size effect, Breit the frequency-independent transverse photon interaction, $\omega$TP the frequency-dependent transverse photon interaction calculated with DHF wave functions, MS the mass shift, QED the QED estimation based on screened-hydrogenic approximation, SDc the correlations energy arising from single and double electron exchanges, and HO the systematic effect summarising all other uncounted terms. The values of the DHF$_0$, Breit and $\omega$TP terms are are basis dependent. Such a basis-dependency is lifted after taking into account all correlation effects. Thus, their uncertainties are effectively accounted for in the uncertainties of SDc and HO terms.  All values are shown in units of eV with two decimal digits. However, in the main text, the total binding energy is rounded up to an integer value, i.e., $167847(5)$~eV.}
\label{tab:bindingE}
\end{center}
\end{table}
\begin{table}[H]
    \setlength{\tabcolsep}{2pt}
\begin{center}
\begin{tabularx}{160mm}{rlll}
\hline\hline
Contribution      &                                        & $^{118}$Sn$^{49+}$ &  Refs. \\ \hline 
Dirac value       &                                         & \phantom{-}1.908 079 205 3(1)  &  \cite{breit_magnetic_1928,beier_g_j_2000} \\ 
Finite size       &                                         & \phantom{-}0.000 014 489 4(110)(211)  & TW, methods \\ 
& & &\hspace{5pt}\cite{Yerokhin2013,angeli_table_2013,GRASP2018, Zatorski12,Karshenboim2018,michel_nonperturbative_2019,Nefiodov02, cakir_improved_2020,jentschura_mass_2006} \\ 
1-loop QED        & $(Z \alpha)^0$                          & \phantom{-}0.002 322 819 5  & \cite{Schwinger1948, tiesinga_codata_2021} \\ 
                  & SE binding                              & \phantom{-}0.000 182 170 1   & \cite{Yerokhin04} \\ 
                  & SE-FS                                   & -0.000 000 159 4(21)   &  \cite{Yerokhin2013,Yerokhin04,beier_g_j_2000} \\ 
                  & VP-EL, Uehling                          & -0.000 035 203 4   &  \cite{beier_g_j_2000, Karshenboim2001,peskin_introduction_2018,Karshenboim2005} \\ 
                  & VP-EL, Uehling FS                       & \phantom{-}0.000 000 113 7(5)   & \cite{beier_g_j_2000,Yerokhin2013,Karshenboim2005} \\ 
                  & VP-EL, WK                               & \phantom{-}0.000 000 658 4 &  \cite{beier_g_j_2000} \\ 
                  & VP-ML                                   & \phantom{-}0.000 001 225 0(40)   & \cite{Lee2005,beier_g_j_2000} \\ 
                  & muonic VP                               & -0.000 000 002 7(14)  &  TW, methods \\
                  &                                         &   &\hspace{5pt}\cite{Belov2016,Karshenboim2005} \\ 
2-loop QED        & $(Z \alpha)^0$                          & -0.000 003 544 6   &  \cite{Petermann1957,Sommerfield1958,tiesinga_codata_2021} \\ 
                  & $(Z \alpha)^2$                          & -0.000 000 078 6    & \cite{Grotch1971,Close1971,Faustov1970,czarnecki_anomalous_2000} \\ 
                  & $(Z \alpha)^4$                          & -0.000 001 123 8   &  \cite{Pachucki05,Czarnecki16} \\ 
                  & $(Z \alpha)^5$                          & \phantom{-}0.000 000 211 0(215)  &  \cite{czarnecki_two-loop_2018,czarnecki_logarithmically_2020} \\ 
                  & $(Z \alpha)^{5+}$ S(VP)E, SEVP, VPVP    & \phantom{-}0.000 000 286 5(53) &  \cite{yerokhin_two-loop_2013,debierre_two-loop_2021} \\ 
                  & $(Z \alpha)^{6+}$ SESE                  & \phantom{-}0.000 000 000 0(2968) &  \cite{czarnecki_two-loop_2018,Pachucki05} \\ 
$\geq$ 3-loop QED & $(Z \alpha)^0$                          & \phantom{-}0.000 000 029 5      & \cite{tiesinga_codata_2021,Laporta1996,Kinoshita2007,Kinoshita2012,Laporta2017,Kinoshita2017} \\ 
                  & binding                                 &\phantom{-}0.000 000 000 7(105)  &   \cite{Grotch1971,Close1971,Faustov1970,czarnecki_anomalous_2000,Pachucki05} \\ 
Recoil            & $\frac{m}{M}$, all-order $Z \alpha$     & \phantom{-}0.000 000 726 8     &  \cite{Shabaev02,beier_g_j_2000,GRASP2018,kramida_nist_2021} \\ 
                  & h.o $\frac{m}{M}$                       & -0.000 000 000 1  &  \cite{Grotch1971,Close1971,Faustov1970,Pachucki08} \\ 
                  & rad-rec                                 & -0.000 000 000 5(4)   & \cite{Grotch1971,Close1971,Faustov1970,beier_g_j_2000,Pachucki08} \\ 
Hadronic          & Uehling                                 & -0.000 000 001 8(14)  &  TW, methods \\
& & & \hspace{5pt}\cite{Belov2016,Friar1999,Eidelman1995,Borie1981} \\
\hline
Sum & & \phantom{-}1.910 561 821 0(2988)  & \\
\hline
Experiment & & \phantom{-}1.910 562 059 0(9) & \\\hline\hline
\end{tabularx} 
\caption{\textbf{Extended Data Table 3: Theoretical \textit{g}-factor} contributions to the bound-electron $g_e$ factor in $^{118}$Sn$^{49+}$. 
“TW” refers to results calculated in this work.}
\label{tab:g}
\end{center}
\end{table}
\begin{figure}[H]  
    \centering
    \includegraphics{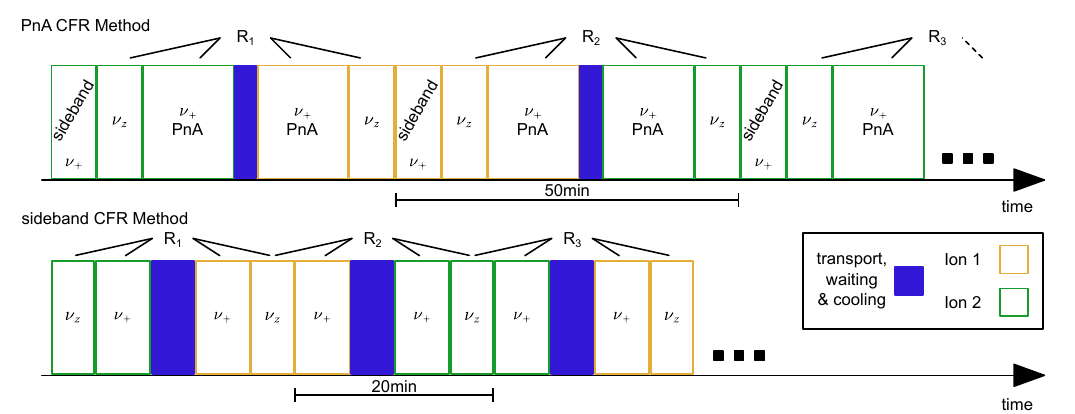}
    \caption{\textbf{Extended Data Fig. 4: CFR Measurement Scheme}
        Two different measurement schemes have been employed for the mass determination.
        In the PnA scheme each ratio has their own axial frequency measurement.
        The $\nu_+$ sideband detection is used in the PnA cycles in order to avoid wrong phase unwrapping.
        Apart from this, it is not used in the CFR evaluation.
        The sideband CFR method consists of interleaved 'double dip' measurements, with axial dips in between.
        As the ions are switched every cycle, each axial dip measurement is used for both neighbouring $\nu_+$ measurements.
    }
    \label{fig:scheme}
\end{figure}
\begin{table}[H]
    \setlength{\tabcolsep}{2pt}
\begin{center}
\begin{tabularx}{135mm}{rlll}
\hline\hline
Observable & System & BS-QED contrb. (theo) &  Experimental Accuracy \\ \hline \hline 

Electronic Lamb shift   & & & \\

 & $\mathrm U^{91+}$ \cite{gumberidze_quantum_2005} & \SI{265.07(53)}{\eV} \cite{yerokhin_lamb_2015} & \SI{4.6}{\eV}\\

 & $\mathrm U^{89+}$ \cite{beiersdorfer_measurement_2005} & \SI{41.77(10)}{\eV} \cite{yerokhin_two-photon_2000,yerokhin_energy_2018} & \SI{0.015}{\eV}\\
 
Muonic Lamb shift & & &\\
 & $\mu\mathrm{Mg}$ \cite{beltrami_new_1986} & \SI{178.68(16)}{\eV} & \SI{0.17}{\eV}\\
 & $\mu\mathrm{Si}$ \cite{beltrami_new_1986} & \SI{274.90(22)}{\eV} & \SI{0.27}{\eV}\\
Pionic Lamb shift       &  & &\\
& $\pi\mathrm{N}$ \cite{trassinelli_measurement_2016} & \SI{1.364\,5(19)}{\eV}~\textsuperscript{*} & \SI{0.0050}{\eV}\\
Hyperfine splitting & & & \\
 & \text{Bi}\textsuperscript{80+} \cite{ullmann_high_2017} & \SI{5.1(2)}{\milli\eV} \cite{volotka_test_2012}   & \SI{0.0146}{\milli\eV}\\
 & spec. diff. \cite{ullmann_high_2017} & \SI{0.229\,0(64)}{\milli\eV} & \SI{0.022}{\milli\eV} \\
$g$ factor & & & \\
 & \textsuperscript{12}\text{C}\textsuperscript{5+} \cite{sturm_high-precision_2014} & \SI{0.8433762(98)e-6}{}~\textsuperscript{\textdagger,\textdaggerdbl} & \SI{6.7e-11}{} \\ 
 & \textsuperscript{28}\text{Si}\textsuperscript{13+} \cite{sturm_g-factor_2013}     & \SI{5.855954(589)e-6}{}~\textsuperscript{\textdaggerdbl} & \SI{9.9e-11}{} \\
 & \textsuperscript{28}\text{Si}\textsuperscript{11+} \cite{wagner_g_2013} & \SI{0.98134(28)e-6}{} \cite{kosheleva_g_2022} & \SI{1.4e-10}{} \\

 & \textsuperscript{40}\text{Ca}\textsuperscript{17+} \cite{kohler_isotope_2016} & \SI{2.35574(29)e-6}{} \cite{kosheleva_g_2022} & \SI{1.1e-10}{} \\ \hline

 & \textsuperscript{118}\text{Sn}\textsuperscript{49+} & $0.148\,802(300)(11)\times 10^{-3}$ & \SI{9.1e-10}{} (mass unc.)\\ 
 &                    &                             & \SI{8.4e-11}{} ($\Gamma_0$ unc.) \\ \hline\hline
\end{tabularx}
\caption{\textbf{Extended Data Table 4: BS-QED tests} Values for the data points in Fig.~\ref{fig:QEDtest}. 
\textsuperscript{*}In this measurement, the pion mass is extracted from the experiment-theory comparison. 
To account for that in the QED test, the theoretical uncertainty is adjusted to the previous literature value of the charged pion mass.
\textsuperscript{\textdagger}Similarly, this measurement is used to determine the electron mass such that a test of QED can only be as accurate as the accuracy of an independent electron-mass measurement.
Recently, the electron mass has been confirmed with similar accuracy in $\mathrm{HD}^+$ spectroscopy \cite{karr_extraction_2023}, allowing to (basically) use the full experimental precision for a test of QED.
\textsuperscript{\textdaggerdbl}Calculated similar to the \textsuperscript{118}\text{Sn}\textsuperscript{49+} theory $g$ factor as described in the Methods section.
}
\label{tab:QEDtests}
\end{center}
\end{table}


\end{document}